# 3D VIDEO QUALITY METRIC FOR 3D VIDEO COMPRESSION

*Amin Banitalebi-Dehkordi[1], Mahsa T. Pourazad[1,2] and Panos Nasiopoulos[1]*

[1]Department of Electrical and Computer Engineering, University of British Columbia, Canada
[2]TELUS Communications Inc. Canada

## ABSTRACT

As the evolution of multiview display technology is bringing glasses-free 3DTV closer to reality, MPEG and VCEG are preparing an extension to HEVC to encode multiview video content. View synthesis in the current version of the 3D video codec is performed using PSNR as a quality metric measure. In this paper, we propose a full-reference Human-Visual-System based 3D video quality metric to be used in multiview encoding as an alternative to PSNR. Performance of our metric is tested in a 2-view case scenario. The quality of the compressed stereo pair, formed from a decoded view and a synthesized view, is evaluated at the encoder side. The performance is verified through a series of subjective tests and compared with that of PSNR, SSIM, MS-SSIM, VIFp, and VQM metrics. Experimental results showed that our 3D quality metric has the highest correlation with Mean Opinion Scores (MOS) compared to the other tested metrics.

## 1. INTRODUCTION

The appearance of multiview display systems in the consumer market is not far from reality. Multiview content includes multiple video streams captured from the same scene simultaneously. In order to minimize the bandwidth and storage requirements for multiview systems, the huge amount of data of multiview video sequences needs to be efficiently compressed. After the introduction of the High Efficiency Video Coding (HEVC) by the joint video team (JVT) of the ITU-T Video Coding Experts Group (VCEG) and ISO/IEC Moving Picture Experts Group (MPEG), the JCT-3V group (a group under MPEG-VCEG) is now working towards standardizing a HEVC-based 3D video codec (3D-HEVC) [1]. This drive comes from the fact that for the same quality, the HEVC standard has shown 35-50 % bit rate reduction compared to the former advanced video coding standard (H.264/AVC) [2]. Fig. 1 illustrates the structure of the future 3D video transmission pipeline, which is recommended by JCT-3V and is used in the 3D-HEVC codec. As it can be observed, the input of this pipeline includes a limited number of views and their corresponding depth maps. This data is encoded, transmitted and decoded at the receiver side. Then several synthesized views are generated based on the decoded views and their corresponding depth maps to support multiview display systems.

In the process of developing the HEVC-based 3D video coding, the quality of the decoded content as well as the synthesized view needs to be tested. The Common Test Conditions provided by JCT-3V for 3D video coding [3] proposes to measure the average PSNR of the stereo pairs formed from a decoded view and a synthesized view [4]. PSNR and in general 2D video quality metrics do not take into account the depth effect, and the binocular properties of the human visual system (HVS). In the absence of a reliable 3D quality metric, based on a comparison study, the use of other 2D quality metrics including VIFp, VQM, MS-SSIM, and SSIM instead of PSNR has been recently proposed to JCT-3V based on a comparison study presented in [5].

It is obvious that the above metrics do not really represent the way the human visual system works especially in fusing 3D information and thus they are not the best objective quality metrics for 3D quality of experience. In an earlier study, we have designed a human visual-system-

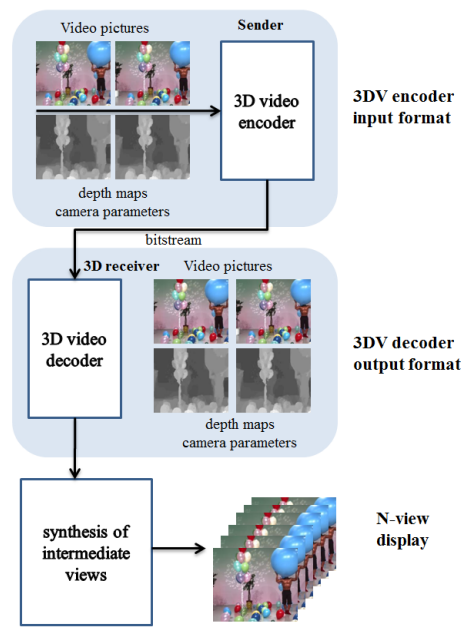

Fig. 1. Overview of the system structure and the data format for the transmission of 3D video

This work was supported by the Natural Sciences and Engineering Research Council of Canada (NSERC) and the Institute for Computing, Information and Cognitive Systems (ICICS) at UBC.

based quality metric for 3D videos known as HV3D [6]. One of the applications of the HV3D metric is to evaluate the quality of compressed 3D videos. In this paper, we test the performance of the HV3D quality metric in the context of the future multiview transmission pipeline (JCT-3V) as shown in Fig. 1. Here, several multiview streams and depth maps are coded with different bitrates using the current version of 3D-HEVC codec [7] and additional views are synthesized at the receiver end. This is a special case of distortions, as compression distortions are combined with distortions generated in the synthesized views. The performance of our proposed metric is compared against subjective test as well as the other metrics suggested in [5].

The rest of this paper is organized as follows: Section 2 provides a brief overview of our HV3D quality metric, Section 3 elaborates on the performance evaluation, and Section 4 concludes the paper.

## 2. HUMAN-VISUAL-SYSTEM-BASED 3D (HV3D) QUALITY METRIC

The HV3D quality metric has been designed having the human visual 3D perception in mind. It takes into account the quality of the individual right and left views, the quality of the cyclopean view (the fusion of the right and left view, what the viewer perceives), as well as the quality of the depth information, as follows [6]:

$$HV3D = w_1 Q_{R'} + w_1 Q_{L'} + w_2 Q_{R'L'} + w_3 Q_{D'},$$

$$w_1 Q_{R'} = [w_1 VIF(Y_R, Y_{R'}) + w_4 VIF(U_R, U_{R'}) + w_4 VIF(V_R, V_{R'}),$$

$$w_1 Q_{L'} = w_1 VIF(Y_L, Y_{L'}) + w_4 VIF(U_L, U_{L'}) + w_4 VIF(V_L, V_{L'}),$$

$$w_2 Q_{R'L'} = w_2 VIF(D, D')^\beta \cdot \sum_{i=1}^{N} \frac{SSIM(IDCT(XC_i), IDCT(XC'_i))}{N},$$

$$w_3 Q_{D'} = w_3 VIF(D, D')^\beta \cdot \sum_{i=1}^{N} \frac{\sigma^2_{d_i}}{N \cdot \max(\sigma^2_{d_j} \mid j = 1,2,...,N)}],$$

(1)

where $Q_{R'}$ and $Q_{L'}$ are the quality of the distorted right and left views compared to the reference views respectively, $Q_{R'L'}$ is the quality of the cyclopean view, $Q_{D'}$ is the quality of the depth information of distorted views, $Y_R$ and $Y_{R'}$ are luma information of the reference and distorted right views respectively (similarly, indexes $L$ and $L'$ denote the left view contents), $U_R$ and $V_R$ are the chroma information of the reference right-view, $U_{R'}$ and $V_{R'}$ are the chroma information of the distorted right-view, $XC_i$ is the cyclopean-view model for the $i$th matching block pair in the reference 3D view, $XC'_i$ is the cyclopean-view model for the $i^{th}$ matching block pair in the distorted 3D view, IDCT stands for inverse 2D discrete cosine transform, $D$ is the depth map of the reference 3D view, $D'$ is the depth map of the distorted 3D view, $N$ is the total number of blocks in the each view, $\beta$ is a constant, $\sigma^2_{d_i}$ is the variance of block $i$ in the depth map of the 3D reference view, SSIM is the structural similarity, VIF is the visual information fidelity index, and $w_1$, $w_2$, $w_3$ and $w_4$ are weighting constants. The weighting constants are chosen so that the quality components used in our method are given different importance in order to lead to the best possible results.

As the equation (1) illustrates, the quality of each view is measured by combining the VIF index of luma and chroma components of the view. To measure the quality of the cyclopean view, it is first modeled by fusing the low frequency information of the left and right (the human visual system is more sensitive to the low frequencies) and intensifying the frequencies that are more important according to the human visual perception system using a Contrast Sensitivity Function (CSF). Then, the quality of the cyclopean view is measured by calculating the SSIM of the distorted modeled cyclopean view. Moreover, the effect of geometrical distortions such as vertical parallax on the cyclopean view is taken into account by measuring the quality of the estimated depth maps from the distorted views (refer to [6] for more details). The quality of the depth information is measured based on the VIF index of the depth maps estimated from the distorted views with the consideration of the size of the display and the distance of the viewer from the screen (see [6] for more details).

The maximum HV3D index value occurs when the modified video is identical to the reference video. To ensure that this maximum is equal to 1, the equation (1) is divided by its maximum value as follows:

$$\hat{HV3D} = \frac{HV3D}{HV3D_{max}}$$

(2)

where $HV3D_{max}$ is the maximum value of HV3D. Since the maximum possible value of SSIM and VIF in equation (1) is unity, $HV3D_{max}$ is equal to:

$$HV3D_{max} = 2w_1 + 4w_4 + w_2 + w_3 \cdot \sum_{i=1}^{N} \frac{\sigma^2_{d_i}}{N \max\{\sigma^2_{d_j} \mid j = 1,2,...,N\}}$$

(3)

## 3. PERFORMANCE EVALUATION

In this paper, the performance of the HV3D metric is tested in the context defined by JCT-3V for the future multiview transmission pipeline, where few multiview streams with their corresponding depth maps are coded and transmitted, and at the receiver side the compressed streams are decoded and are used to generate several synthesized views to support the different multiview display systems [1]. To evaluate the performance of HV3D, subjective tests were performed and the correlation between the mean opinion scores (MOS) and HV3D was calculated. In our experiments the quality of the stereo pairs formed from the decoded left view and a synthesized view is tested with respect to the original pair (see Fig. 2). Note that if the original intermediate view does not exist, this view is synthesized based on the original views and their corresponding depth maps.

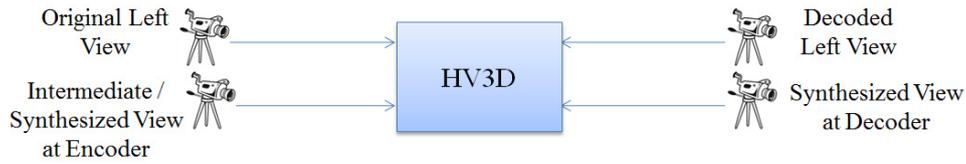

Fig. 2. Application of HV3D quality metric in the 2-view scenario

Table 1: Input views and stereo pairs for 2-view test scenario

| Seq. | Test sequence | Test class | Input views | View to synthesize | Stereo |
|---|---|---|---|---|---|
| S01 | Poznan_Hall2 | A | 7-6 | 6.5 | 6.5-6 |
| S02 | Poznan_Street | A | 4-3 | 3.5 | 3.5-3 |
| S03 | Undo_Dancer | A | 2-5 | 3 | 3-5 |
| S04 | GT_Fly | A | 5-2 | 4 | 4-2 |
| S05 | Kendo | C | 3-5 | 4 | 4-5 |
| S06 | Balloons | C | 3-5 | 4 | 4-5 |
| S07 | Lovebird1 | C | 6-8 | 7 | 7-8 |
| S08 | Newspaper | C | 4-6 | 5 | 5-6 |

The test sequences are selected from the dataset provided by MPEG for the call for proposals (CFP) on 3D video coding (3DV) [8]. Table 1 shows the test videos and synthesized views used in our experiment. For each test sequence, four different bitrates were studied. Specifically, the quantization parameter (QP) was set to 25, 30, 35, and 40. The test sequences were coded using the HEVC-based 3DV codec (3D-HTM 4.0) [7]. The viewing conditions for the subjective tests were set according to the ITU-R Recommendation BT.500-11 [9]. Eighteen observers participated in the subjective tests, with age ranging from 21 to 29 years old. All subjects had none to marginal 3D image and video viewing experience. They were all screened for color and visual acuity (using the Ishihara and Snellen charts), and for stereo vision (Randot test – graded circle test 100 seconds of arc). The evaluation was performed using a 46" Full HD Hyundai 3D TV (Model: S465D) with passive glasses. The TV settings were as follows: brightness: 80, contrast: 80, color: 50, R: 70, G: 45, B: 30, as recommended by MPEG for 3DV CFP.

After a short training session, the viewers were shown the compressed and the reference stereo pair sequences in a random order. This would allow the viewer to watch the reference and the compressed versions of the same sequence consecutively without knowing which video is the reference one. Between the test videos, a four-second gray interval was provided to allow the viewers to rate the perceptual quality of the content and relax their eyes before watching the next video. Here, the perceptual quality reflects whether the displayed scene looks pleasant in general. In particular, subjects were asked to rate a combination of "naturalness", "depth impression" and "comfort" as suggested by [10]. There were 10 quality levels (1-10) for ranking the videos, where score 10 indicated the highest quality and 1 indicated the lowest quality. After collecting the experimental results, outliers were removed (there were three outliers) and then the mean opinion scores from viewers were calculated.

The performance of HV3D was compared with the suggested 2D quality metrics by [5]. Fig. 3 shows the logistic fitting curve associated with each of the quality metrics. The vertical axis shows the MOS and the horizontal axis represents quality metric measures. As it can be observed, the values resulting from HV3D have the highest correlation with the subjective test results.

Correlation of each quality metric with MOS is calculated in terms of Pearson and Spearman Correlation Coefficients (PCC and SCC). Table 2 shows the correlation ratios for different metrics. We observe that HV3D achieves the highest correlation with the perceived quality compared to other metrics.

## 4. CONCLUSION

In this paper the performance of our HV3D quality metric for multiview applications was verified through a series of subjective tests and was compared with other suggested quality metrics. Experimental results illustrate that HV3D has the highest correlation with the perceived quality compared to PSNR, SSIM, MS-SSIM, VIFp, and VQM. The superior performance of this metric suggests that it is a perfect candidate for evaluating the quality of compressed videos within the framework of the 3D-HEVC standardization efforts.

Table 2: Pearson and Spearman Correlation Coefficients for different quality metrics

| Metric | Spearman rank correlation coefficient (SCC) | Pearson linear correlation coefficient (PCC) |
|---|---|---|
| PSNR | 0.6357 | 0.6554 |
| SSIM | 0.6709 | 0.7034 |
| VQM | 0.6845 | 0.6805 |
| VIFp | 0.7188 | 0.7475 |
| MS-SSIM | 0.8033 | 0.7916 |
| **HV3D** | **0.8646** | **0.8566** |

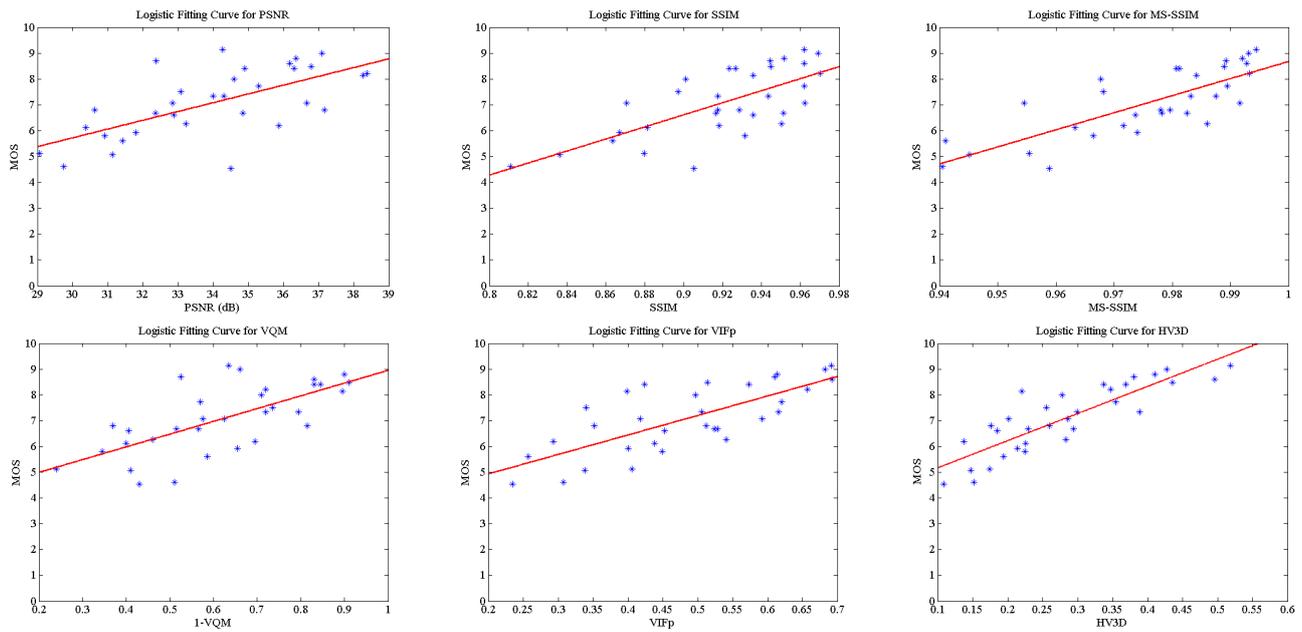

Fig. 3. Comparison of the subjective results (MOS) with six different quality metrics